# AI Identification: An Integrated Framework for Sustainable Governance in Digital Enterprises

Di Kevin Gao: California State University Sacramento
Jingdao Chen: Mississippi State University
Shahram Rahimi: University of Alabama

**Abstract**

As artificial intelligence (AI) systems grow more powerful, autonomous, and embedded in critical infrastructure, their identification and traceability become foundational to regulatory oversight and sustainable digital governance. In digitally transformed enterprises, long-term sustainability depends on transparent, accountable, and lifecycle-governed AI systems, all of which require verifiable identity. This study proposes a conceptual and architectural framework for AI identification, combining technical and governance mechanisms to support lifecycle accountability. The framework integrates five components: model fingerprinting, cryptographic hashing, blockchain-based registration, zero-knowledge proof (ZKP)-based proof of possession, and post-deployment structural change screening. We introduce a dual-layer identifier, consisting of a machine-verifiable primary hash and a human-readable secondary identifier, anchored in a tamper-resistant registry. Identity validation is supported by selective ZKP-based verification at governance-defined checkpoints, while post-deployment changes are monitored using Lempel–Ziv Jaccard Distance (LZJD) as a governance-oriented screening signal rather than a semantic performance metric. The framework establishes an enforceable and transparent identity infrastructure that enables continuity, auditability, and policy-aligned oversight across AI system lifecycles. By embedding AI identification within enterprise architecture and governance processes, the proposed approach supports sustainable innovation, strengthens institutional accountability, and provides a foundation for selective, policy-defined verification during digital transformation.

## I. Introduction

AI has demonstrated formidable capabilities across domains in healthcare, legal reasoning, and even pandemic responses [1]– [7] and beyond. These developments are fueled by advances in machine learning, greater data accessibility, an exponential rise in computational power [8], and an unprecedented consumption of energy resources [9]. Despite the accelerating sophistication of artificial intelligence systems, particularly multimodal large language models (MLLMs) capable of human-like reasoning and conversation, their regulatory traceability remains profoundly underdeveloped [10]. As these systems increasingly mediate digital, economic, and even civic interactions [11], a foundational challenge emerges: How can society establish a reliable, cryptographically sound, and legally operable identity for artificial intelligence systems? How can one verify that a given AI is what it claims to be?

The question is not merely technical. In the absence of standardized AI identifiers, users face epistemic asymmetries; regulators encounter enforcement bottleneck; and firms, paradoxically, operate without credible means to demonstrate compliance. Indeed, as Gao et al. have noted, to the general public, AI remains an amorphous force rather than a verifiable product [12]. While some jurisdictions, notably the European Union and China [13], [14], have introduced mandatory AI registration requirements for certain AI models, the underlying architectures resemble conventional metadata collection schemes, insufficiently equipped to handle the opacity, adaptivity, and non-determinism that characterize advanced AI systems. We conducted an extensive review of the existing literature and found remarkably few studies that directly address this pressing need.

In addition, the absence of reliable AI identification mechanisms now poses a growing governance challenge for digitally transforming enterprises. As organizations adopt AI at scale, governance-enforceable traceability, characterized by transparency, accountability, lifecycle oversight, and responsible resource use, requires that AI systems be consistently identifiable across architectures, platforms, and operational contexts [15]. Without verifiable identity, digital governance frameworks, enterprise architecture processes, and risk-based regulatory models lack the foundational visibility needed to manage AI safely and sustainably [16].

This paper addresses this critical gap by proposing an integrated framework by using an AI model's weight configuration as its fingerprint which effectively binds the system's behavior to its identity. It then recommends a cryptographic hashing of this fingerprint to yield a primary identifier suitable for machine verification. It further proposes a coding scheme for a secondary ID suitable for human readers. To ensure auditability, this paper recommends that AI IDs be immutably stored on a blockchain registry, while deploying zero-knowledge proof(ZKP)-based checkpoint verification protocols to prove AI identities without revealing sensitive internals. Post-deployment consistency is monitored through weight similarity scoring, using Lempel-Ziv Jaccard Distance (LZJD) to detect model drift and trigger re-registration where needed.

While grounded in technical implementation, the proposal is also policy-aware. It avoids imposing excessive disclosure burdens on firms, leverages tamper-resistant infrastructure to support public oversight, and accommodates dynamic AI systems without sacrificing accountability. It is important to note that the framework can be understood as a bridge between the cryptographic logic of software integrity and the institutional logic of regulatory legitimacy. More importantly, by enabling traceability and lifecycle verification, AI identification becomes an essential capability for sustainable digital governance, supporting enterprise architecture oversight, risk-informed deployment, and responsible digital transformation.

This article therefore contributes not only to the technical discourse on AI provenance but also to the broader challenge of building sustainable, accountable, and future-ready digital governance systems. It advances the academic discourse on AI identification by proposing an end-to-end, integrated framework that encompasses model fingerprint generation, cryptographic hashing, identity assignment, and post-deployment drift detection. Although a growing body of research addresses AI transparency, accountability, and governance, the problem of persistent system-level identification remains under-specified. Existing approaches typically focus on documentation (e.g., Model Cards and System Cards), behavioral inference (e.g., output-based fingerprinting or probing), or organizational governance frameworks (e.g., risk management standards and audit processes). These tools improve visibility and oversight, but they do not establish a verifiable, cryptographically anchored identity that persists across deployments, versions, and operational contexts. This work defines AI identification as the assignment and verification of a lifecycle-bound, system-level identity tied directly to a model's internal configuration, rather than to its documentation, observed behavior, or organizational process. Identification discussed here is distinct from authentication of individual interactions, inference outputs, or users, and from post hoc auditing or behavioral analysis. The research gap addressed here is therefore not how to describe, evaluate, or govern AI systems in general, but how to ensure that a specific deployed AI instance can be reliably recognized as the same system that was registered, evaluated, and approved earlier in its lifecycle. To ensure analytical clarity and avoid overreach, we need to define scopes and boundaries. This study is explicitly positioned as a conceptual, architectural, and governance-oriented contribution rather than an empirical systems evaluation. While this work addresses AI identification in the context of sustainable digital governance, it does not attempt to solve sector-specific compliance, nor provide energy-consumption benchmarking for individual AI models. Instead, the contribution lies in proposing a generalizable and technology-agnostic identity infrastructure that strengthens lifecycle accountability across jurisdictions. Its validity is assessed in terms of governance feasibility, architectural coherence, and regulatory operability rather than performance benchmarking. Sustainability in this context is defined not environmentally alone, but as long-term institutional resilience: the ability of digital governance systems to support transparent provenance, reduce redundant model retraining and recertification, enable cross-platform interoperability, and maintain regulatory continuity amid rapid model evolution. These boundaries clarify the analytical scope of the paper and avoid overstating claims beyond the architectural and governance dimensions addressed here.

## II. Background

As artificial intelligence systems, particularly multimodal large language models (MLLM), advance in sophistication and become seamlessly embedded in digital and physical environments, the frequency and complexity of human–AI interaction continue to escalate. A pressing concern accompanying this evolution is the increasing difficulty in discerning whether one is engaging with a human agent or an autonomous system [17]. In light of this development, a foundational step toward safe and accountable AI deployment is the formal identification of AI systems. This foundational requirement is increasingly recognized as a prerequisite for persistent system identity, where transparency, accountability, and lifecycle oversight are essential to managing AI responsibly within complex socio-technical environments.

In this paper, AI identification refers to the assignment and verification of a persistent, system-level identity bound to an AI model's internal configuration, rather than to users, sessions, or inference outputs. This identity enables verifiable provenance, lifecycle traceability, and governance-enforceable recognition of a specific AI system across deployment contexts. By providing a stable reference point for oversight, AI identification supports regulatory traceability, institutional accountability, and trust in AI-enabled systems.

The academic literature has begun to acknowledge the urgency of this issue. Gao and Haverly et al. identified AI identification as a neglected but vital research domain. They recommend the introduction of standardized AI identifiers to improve recognition, ensure consistent categorization, and align user expectations. Specifically, the authors advocate for a central registry, potentially supported by blockchain technology, to operationalize this goal [18]. This line of research parallels broader calls in digital governance for verifiable system provenance and lifecycle management, both of which are essential for sustainable innovation in AI-enabled enterprises.

Policymakers are beginning to respond. The European Union's AI Act mandates the creation of a database for high-risk AI systems [19], though the associated technical standards are still under development. Similarly, China's 2021 regulation on recommendation algorithms established a mandatory registration regime for algorithmic systems [14]. While countries such as South Korea, Singapore, the United Arab Emirates, and Japan have not enacted binding legal requirements, they have issued non-binding guidelines encouraging transparency and voluntary disclosures. In the United States, comparable efforts have emerged at the subnational level. In October 2025, California enacted legislation requiring conspicuous user notification when interacting with companion chatbots [20]. Collectively, these developments reflect growing legislative interest in AI transparency and traceable identification. They also highlight a global recognition that sustainable digital governance depends on mechanisms that ensure AI systems are knowable, accountable, and governable throughout their lifecycle [21].

## III. AI Identification as a Governance-Enforceable Traceability Capability

Artificial intelligence systems increasingly form the operational backbone of digital enterprises, public infrastructures, and cross-organizational data ecosystems. As AI systems grow in complexity, their traceability and identity assurance become essential components of sustainable digital governance. Sustainable digital governance requires mechanisms that ensure transparency, accountability, lifecycle oversight, and regulatory compliance without imposing excessive operational burdens. Recent governance research emphasizes that transparency and accountability are foundational pillars for responsible AI systems and that legal and ethical frameworks must be balanced with operational feasibility to uphold societal wellbeing [22]. The proposed AI ID framework directly supports these objectives. Cryptographic fingerprints, blockchain-based identity registration, and zero-knowledge proofs allow AI systems to be verifiably authentic while minimizing energy-intensive duplication of model training and reducing risks associated with the deployment of unregistered or tampered AI systems. As a result, AI identification emerges as a governance capability that simultaneously enhances safety, reduces environmental externalities, and fosters longterm sustainability within digital ecosystems.

In this work, sustainability is understood not solely in environmental terms, but as long-term institutional and governance resilience in digitally transformed enterprises. The proposed AI identification framework contributes to this form of sustainability by enabling persistent identity, lifecycle traceability, and verifiable reuse of AI systems across organizational and regulatory contexts. By reducing uncertainty surrounding model provenance and compliance status, the framework lowers incentives for redundant model retraining, repeated certification, and duplicative audits, all of which are resource-intensive activities. Accordingly, AI identification serves as enabling infrastructure that supports more efficient allocation of computational, organizational, and regulatory resources over time.

Beyond regulatory compliance, AI identification creates economic incentives for adoption by reducing duplicated governance effort across organizations and lifecycle stages. When a verifiable identity persists across deployment contexts, firms can avoid repeated certification, redundant audits, and unnecessary retraining cycles, all of which impose material computational and organizational costs. For our discussion, AI identification functions not only as a compliance mechanism, but also as efficiency-enhancing infrastructure for digitally transforming enterprises.

While this paper does not present empirical measurements, the sustainability contribution of the framework can be evaluated using measurable indicators, including:

- **Reduction in redundant model retraining:** the proportion of AI deployments that reuse an existing registered identity rather than initiating new training or fine-tuning cycles.
- **Compliance and audit efficiency:** the reduction in time and organizational effort required to verify AI provenance, registration status, and version lineage during regulatory or internal audits.
- **Lifecycle accountability coverage:** the share of deployed AI systems within an organization or sector that maintain continuous identity traceability from initial registration through deployment, update, and retirement.

## IV. Embedding AI Identification into the AIDAF FSAO Digital Transformation Process

The Adaptive Integrated Digital Architecture Framework (AIDAF) and its FSAO process as indicated in Figure 1, comprising Communication, Integration, Adaptation, and Digitalization phases, offer a structured enterprise architecture approach for digital transformation [23], [24]. AI Identification can be mapped directly into this governance cycle as follows:

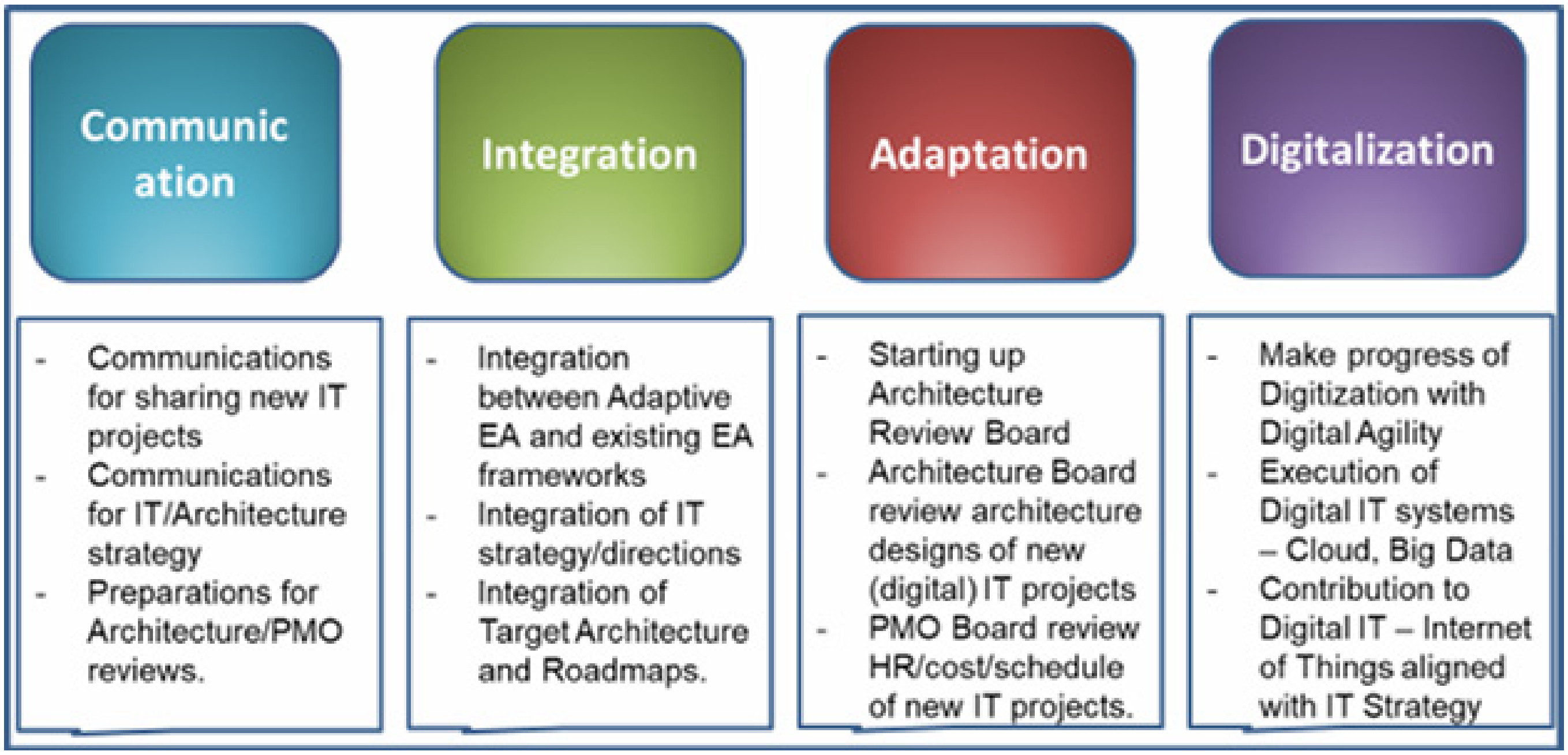

Fig. 1. FSAO Digital Transformation Process.

1) Communication Phase: Organizations broadcast new AI projects, model updates, and lifecycle changes. AI IDs provide consistent metadata, enabling clear communication of model provenance, versioning, and intended risk categories.
2) Integration Phase: AI IDs become integrated into enterprise architecture guidelines, IT governance policies, data governance frameworks, and risk management procedures. The identifiers serve as reference artifacts across business, application, data, and technology architecture layers.
3) Adaptation Phase: Within the Adaptation Phase, technical modifications such as fine-tuning, pruning, or parameter updates result in a change to the model's structural fingerprint, triggering recomputation of the model-level hash and, where applicable, a new namespaced registry commitment. This update propagates through the identification pipeline, requiring re-registration or governance review depending on predefined thresholds. In this way, AI Identification introduces a concrete technical artifact into the adaptation process, directly linking enterprise transformation activities to verifiable system identity.
4) Digitalization Phase: As digital systems go live, checkpoint verification using ZKP proofs ensures deployed AI remains authentic and within acceptable drift thresholds. Drift triggers re-registration, ensuring sustainable lifecycle governance.

Through this mapping, AI Identification becomes embedded directly into the enterprise digital transformation pipeline, aligning with AIDAF's governance philosophy and strengthening the link between technical controls and strategic digital objectives.

The AI identification framework proposed in this work is not intended to replace existing enterprise architecture or governance frameworks such as AIDAF or FSAO. Rather, it is designed to operate as a complementary technical capability within such frameworks. While enterprise and governance models define processes, roles, and oversight structures, they generally assume the existence of a stable and verifiable system identity. AI identification addresses this assumption by providing a lifecycle-bound, cryptographically anchored identity layer that can be integrated into existing architectural and governance workflows without altering their normative structure.

At the level of enterprise architecture implementation, the proposed framework reduces resource-intensive duplication of model certification by allowing a single cryptographic identity to persist across architectural environments, provided drift remains within approved thresholds. It further reduces carbon-intensive retraining workloads by supporting identity continuity for models that undergo minor adaptations, aligning with recent findings that repetitive fine-tuning cycles are a non-trivial driver of AI-related emissions [25]. The framework therefore contributes to sustainability not by reducing compute demand directly, but by creating the institutional conditions under which compute resources, regulatory attention, and model audit lifecycles are allocated more efficiently over time.

## V. Key Challenges in AI Identification

Designing a robust and future-proof identification and registration regime for AI will require confronting a range of complex and interdependent challenges, both technical and regulatory [26], [27]. Because these challenges directly shape the transparency, accountability, and lifecycle assurance required for sustainable digital governance, resolving them is essential for organizations and regulators engaged in responsible digital transformation.

- Uniqueness: What intrinsic features distinguish one AI model from another? This question relates to the creation of digital "fingerprints" based on model architecture, weights, or behavior.
- Conversion: How can these fingerprints be converted into secure, tamper-proof identifiers? Here, cryptographic hashing and the generation of both machine-readable and human-readable IDs play a critical role.
- Storage: What is the optimal storage mechanism for these identifiers? Here, a decentralized, tamper-resistant blockchain registry offers promising advantages.
- Verification: How can an AI's identity be verified at governance-defined checkpoints during deployment and operation? Based on our research, zero-knowledge proof (ZKP) protocols enable privacy-preserving identity verification without disclosing sensitive model details, while remaining compatible with current scalability constraints.
- Integrity: How can model drift or post-deployment tampering be detected in a scalable and governance-enforceable manner? Here, techniques such as Lempel–Ziv Jaccard Distance (LZJD) scoring may provide a practical screening signal for structural divergence to support integrity attestation decisions.

## VI. State of the Arts and Research Gaps

Despite growing interest in AI governance, few academic works offer a comprehensive technical framework for AI identification that enables persistent, system-level provenance and governance-enforceable identity continuity.

OECD's framework for the classification of AI systems provides valuable risk-based taxonomies [28]. However, it does not define mechanisms for persistent system-level identification, version continuity, or cross-context traceability, leaving a critical implementation gap between classification and enforceable oversight.

Beyond these classification efforts, several adjacent approaches are sometimes cited as mechanisms for AI provenance or accountability. However, these approaches do not address the problem of persistent system-level identification. For example, watermarking and content signatures embed signals within AI-generated artifacts to detect whether an output was machine-produced, but they do not identify which model produced the output nor provide a cryptographic anchor to the model's internal configuration. Likewise, Model Cards and System Cards primarily support transparency and documentation rather than identity: they describe a model's intended use, training data, and known limitations, but do not constitute a machine-verifiable identifier or enable lifecycle traceability. Similarly, governance frameworks such as the NIST AI RMF and ISO/IEC 42001 [29] offer process-and policy-level structures for safe deployment, yet lack mechanisms for binding an AI system to a verifiable identity at the technical layer. Finally, behavior-based fingerprinting and model probing strategies can infer identity characteristics from observed outputs, but these methods remain probabilistic, are not cryptographically secured, and may not guarantee consistency across fine-tuning events or deployment environments. Traceability and auditing mechanisms track AI usage, risk, or compliance events over time, but they presuppose the existence of a stable system identity and cannot substitute for identity itself.

Table I summarizes these distinctions and illustrates that while the above mechanisms contribute to AI accountability, none provide a persistent, cryptographically verifiable identity of the model itself, nor do they embed identity within enterprise architecture or lifecycle governance structures as proposed in this work.

Having established that adjacent provenance and governance mechanisms do not provide system-level identification, it is necessary to consider works that directly address the idea of AI identifiers. Among them, Chan et al. introduce the concept of "IDs for AI systems" [30]; however, their formulation is oriented toward instance-level or session-specific tokens rather than the persistent identity of the underlying model. As such, it does not enable continuity across redeployments or model updates.

A second line of work seeks to infer identity from observable behavior. Bhardwaj and Mishra propose a hybrid fingerprinting technique that combines architectural and behavioral traits to approximate the originating model [31]. Pasquini et al.'s LLM Map similarly employs crafted queries to differentiate model versions through output-based signatures [32], while Yang and Wu adopt a black-box probing approach operating under comparable constraints [33]. These methods are valuable for external auditing but remain probabilistic, lack cryptographic guarantees, and may fail under fine-tuning, pruning, or compression, where behavioral characteristics shift despite continuity of lineage.

A third category involves internally generated fingerprints intended for proprietary protection rather than technical governance. Zeng et al.'s HuRef creates human-readable identifiers to support intellectual property claims [34]; however, these identifiers are not designed to persist across heterogeneous deployment environments, nor do they provide verifiable anchoring to the model's weight state. Taken together, these studies indicate growing interest in AI fingerprinting and identification, yet they diverge from the objective pursued in this work: enabling a *persistent, cryptographically anchored identity for the model itself*, resilient to architectural context, model adaptation, and deployment change. This distinction is central to supporting lifecycle traceability and regulatory enforcement, and motivates the framework presented in the following section.

The framework proposed in this work does not aim to evaluate model quality, predict behavior, certify safety, or infer compliance from observed outputs. It does not replace model evaluation, auditing, or risk assessment, nor does it guarantee inference correctness or ethical performance. Instead, AI identification provides the precondition for these activities by ensuring that the system under examination is verifiably the same system that was previously registered, reviewed, or approved. Without such identity continuity, downstream governance mechanisms operate on uncertain or unverifiable referents.

TABLE I
COMPARISON OF ADJACENT APPROACHES AND THEIR RELATION TO SYSTEM-LEVEL AI IDENTIFICATION.

| Concept | Main Applications | Identifies the AI System Itself? |
|---|---|---|
| Watermarking | Embeds signals in generated content to detect if content is AI-produced | No |
| Model Cards | Documentation of a model's purpose, training data, and limitations | No |
| System Cards | High-level documentation for systems, not just models | No |
| NIST AI RMF | Risk management and governance framework | No |
| OECD / ISO 42001 | Policy frameworks and assurance standards | No |
| Content Signatures | Detects AI-generated artifacts or output traces | No |
| Fingerprinting via Behavior (LLM probes) | Infers model identity through behavioral patterns and probing | Partial |
| **This Work** | Weight-based fingerprint + cryptographic hash + ZKP anchor for lifecycle identity | **Yes (persistent and verifiable)** |

## VII. AN INTEGRATED FRAMEWORK TO ADDRESS THE CRITICAL GAPS

In the following sections, we propose an integrated framework for AI identification that combines cryptographic assurance, decentralized governance, and practical enforceability. An overview of the proposed solution is depicted in Figure 2 below. Together, these components provide the technical foundations required for sustainable digital governance, enabling transparent, trustworthy, and lifecycle-verifiable AI deployment at scale.

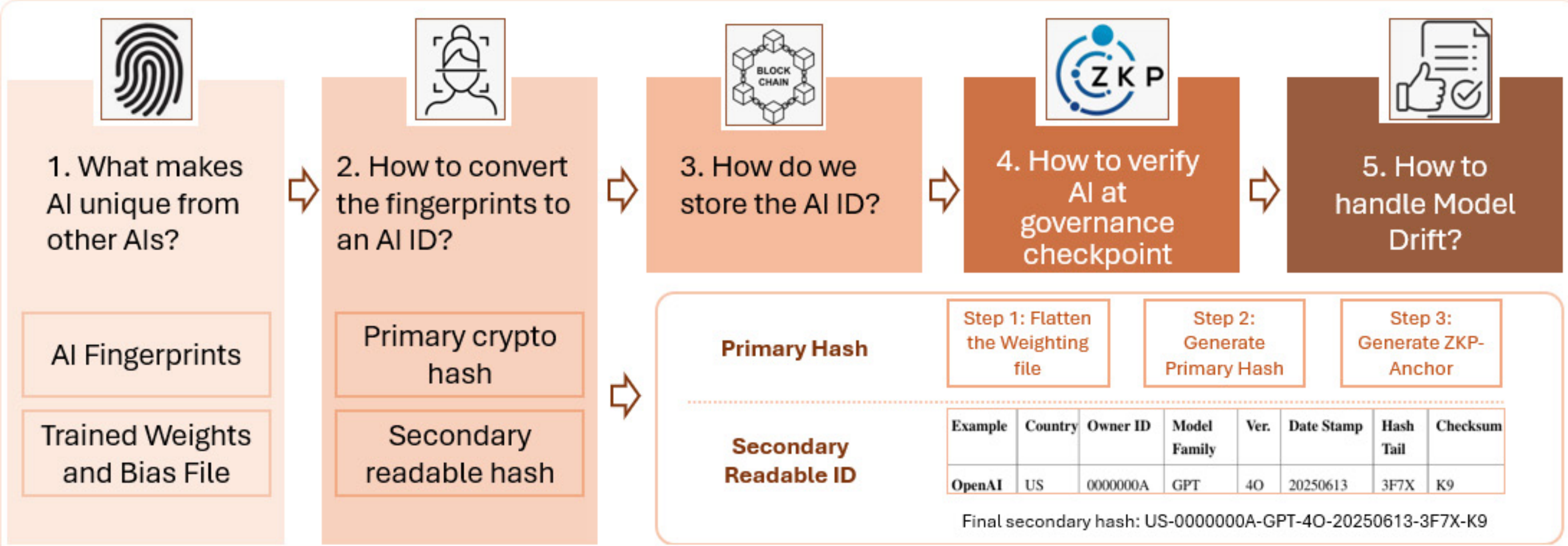

Fig. 2. We propose a unique AI ID generated from model weights via both a machine-readable primary hash and human-readable product label, stored on a tamper-resistant blockchain registry. This enables transparent access and ZKP-based verification at governance checkpoints.

### *A. The AI Fingerprint: Identify AI Models by Their Learned Weights*

Recent development in AI technology has led to a foundational insight that an AI model's weight (and bias) configuration, the complete set of learned parameters governing its behavior, serves as its unique fingerprint. Just as with human fingerprints, distinct weight configurations generally correspond to distinct behavioral profiles under standard evaluation conditions, making learned weights a practical and technically grounded basis for model identification. These weights determine how the AI interprets prompts, processes information, and generates responses. The weight configuration provides a stable and implementation-aligned basis for identifying AI model instances at the version level.

Building on this, we propose that an AI ID should be assigned at the level of each stable weight configuration, with an associated timestamp when that version was fixed or released, plus governing authority information, owner details, and owner-designated product line information. This approach ensures that each AI ID corresponds to a verifiable and reproducible behavior profile, enabling transparency, accountability, and traceability in both deployment and regulation. It avoids overgeneralizing identity across evolving systems while maintaining scalability for large model families.

Beyond minor adjustments, any significant modification to weights, such as those resulting from fine-tuning, retraining, or architectural changes, should trigger the issuance of a new AI ID, equivalent to a software version update. This system captures the dynamic lifecycle progression of AI while ensuring traceability and auditability. When combined with cryptographic hashing and blockchain registration, this version-level ID system provides a secure foundation for provenance verification, checkpoint authentication, and public accountability. As a result, weight-based identification directly supports persistent system identity by enabling consistent lifecycle oversight across rapidly evolving AI systems.

### *B. From Fingerprints to Tamper-Proof Identifiers: Primary and Secondary Hashing*

In the previous subsection, we defined a unique fingerprint for every AI. Conceptually, it is true to say “the weights are the fingerprint,” but that is practically infeasible unless we standardize and compress that fingerprint into something verifiable, comparable, and easy to reference, such as a human-readable AI ID. Additionally, we want to shield the real weights from tampering or theft. Standardization of identifiers is also essential for enterprise architecture governance, as consistent and interoperable AI IDs enable cross-system visibility and responsible digital transformation.

To convert the fingerprint to an AI ID, we recommend a two-layered system which is described in detail in the following section. Before that, we would like to provide a terminology clarification: in this paper, the term *primary hash* refers to the cryptographic commitment computed over a model's deterministically serialized weight configuration, denoted as $H_W$. The term *AI-ID* denotes the externally registered identifier derived from this commitment, for example through namespaced hashing, and recorded in the registry. Where relevant, the term *commitment* is used synonymously with the primary hash in the cryptographic sense.

1) *Primary Hash: Commitment and Namespaced Identity:* The foundation of the proposed identification scheme is a two-stage cryptographic hashing process that binds the AI system’s internal state to a machine-verifiable identity without revealing proprietary model parameters. Let $W$ denote the serialized weight tensor of a deployed model. During registration, the model owner computes a commitment to the full model state by applying a cryptographic hash to the tensor: $H_W =$ SHA-256($W$). This 256-bit digest, represented as a 64-character hexadecimal value, functions as a hiding and binding commitment: it is computationally infeasible to reconstruct $W$ from $H_W$, and any modification to $W$ , even at the level of a single bit, will almost certainly produce a different digest. To namespace identities across organizations and prevent cross-issuer collisions, the authoritative machine-readable identifier (*AI-ID*) is derived by hashing the commitment together with an issuer prefix (e.g., company code or registry code): AI-ID = SHA-256(company_code||$H_W$), where || denotes concatenation. A high-level, non-normative pseudo code is provided in Appendix A. This double-hash design yields a public identifier anchored to both the model and its declaring entity, while preserving the confidentiality of the model’s internal structure. The *AI-ID* is the value recorded in the registry or blockchain ledger, and is the only identifier disclosed to external systems. This two-stage hashing structure also minimizes the likelihood of identifier collision. Even before name spacing, accidental equality between SHA-256 digests of distinct models is already negligible (on the order of $2^{-256}$). By incorporating the company code as part of the input to the second hash, the identifier space is effectively partitioned by issuer; two distinct (company, model) pairs would need to collide simultaneously. Under standard cryptographic assumptions, the probability of such a collision is so small as to be operationally irrelevant for regulatory and governance time horizons. While this does not constitute a proof of global uniqueness in the mathematical sense, it renders duplicate *AI-ID* values practically infeasible. A zero-knowledge proof (ZKP) protocol enables identity verification without exposing model contents or re-evaluating the hash over billions of parameters. Importantly, the ZKP does not re-compute the full SHA-256($W$) hash inside the circuit. While this would provide the strongest possible guarantee of correctness, it is currently infeasible for large-scale models. A model with one billion parameters would require on the order of billions of circuit constraints to represent the hashing function, since current ZKP systems require one or more arithmetic constraints per operation in the hash computation. Even optimistic throughput benchmarks (10,000–50,000 constraints per second on commodity hardware) imply proof generation times measured in days, with memory requirements reaching hundreds of gigabytes. Published ZKML toolchains (e.g., EZKL, Modulus Labs, and Halo2-based prototypes) report practical limits below approximately 10–50 million parameters, significantly below the scale of contemporary foundation models. Accordingly, this work scopes ZKP usage to proving possession of the registered cryptographic commitment rather than performing full in-circuit evaluation of model weights. This design choice reflects current ZKP performance boundaries and avoids reliance on speculative advances in ZKML scalability. Instead, the proof focuses on a tractable target: demonstrating possession of the model that produced the registered commitment. Rather than proving knowledge of the public *AI-ID* itself, which is trivial, the prover demonstrates knowledge of a *preimage* of the commitment: the full tensor $W$ such that $H_W =$ SHA-256($W$) and AI-ID = SHA-256(`company_code`||$H_W$). These equalities are verified inside the ZKP circuit using $W$ as the private witness and *AI-ID* and `company_code` as public inputs. This establishes that the deployed system corresponds to the registered identity without revealing $W$ or executing a full-model hash inside the circuit. The scheme provides proof of possession and integrity rather than proof of model behavior or inference correctness. The framework therefore does not claim to authenticate individual inference outputs, nor to guarantee behavioral equivalence, safety, or correctness at governance checkpoint. In environments where models are accessed exclusively through trusted APIs or centralized platforms, simpler mechanisms such as digital signatures or platform attestations may be sufficient; zero-knowledge proofs are introduced here specifically for contexts in which autonomous verification is required without reliance on a trusted execution intermediary. Because frameworks differ in how weights are serialized (e.g., PyTorch `.pt`, TensorFlow `SavedModel` directory, ONNX `.onnx`), the hashing process requires deterministic encoding of $W$. Each model publisher must standardize its export and serialization routines so that identical models always yield identical commitments. This constraint is essential for cross-organizational reproducibility and for regulators to treat hashed identities as stable references over the lifecycle of versioned and redeployed AI systems.

2) *Secondary Hash:* The secondary hash is a human-readable identifier designed to support transparency, visibility, and traceability for human users. While the primary hash is used for cryptographic verification, the secondary hash serves public-facing applications such as registry display, audit reporting, and deployment approvals. The format presented below represents one standardized option among several possible implementations:

- Country Code: A 2-letter ISO code indicating the country of origin or primary jurisdiction of the AI developer.
- Owner ID: An 8-character alphanumeric identifier assigned to the developing organization.
- Model Family: A 3-character alphanumeric code designated by the owner to represent a family of models.
- Version Code: A 2-character alphanumeric field identifying a specific version within the model family.
- Date Stamp: An 8-digit numeric field formatted as YYYYMMDD, indicating the model's registration or issuance date.
- Hash Tail: A 4-character alphanumeric suffix derived from the last segment of the model's primary hash, providing a deterministic, non-sensitive link to the underlying fingerprint.
- Checksum: A 2-character alphanumeric segment calculated from the concatenation of all preceding fields, using a lightweight hash function to detect transcription or formatting errors.

As an illustration, an example secondary hash for an OpenAI GPT-4o model could be represented as: US-0000000A-GPT4O-20250613-3F7X-K9. Figure 3 provides examples of sample secondary hash structures.
This ID is intended for use in public registries, certification documents, API headers, or other compliance-relevant artifacts. While not cryptographically secure, it provides an interpretable and verifiable label that balances utility with integrity. Human-readable identifiers also strengthen persistent system identity by facilitating transparency for users, auditors, and cross-sector stakeholders.

| Example | Country | Owner ID | Model Family | Ver. | Date Stamp | Hash Tail | Checksum |
|---|---|---|---|---|---|---|---|
| **OpenAI** | US | 0000000A | GPT | 4O | 20250613 | 3F7X | K9 |
| **Copilot** | US | 00000001 | COP | 01 | 20240601 | 7KQ9 | B2 |
| **Claude** | US | 0000000C | CLA | 3B | 20240520 | Z8R1 | X7 |
| **DeepSeek** | CN | 000000D5 | DSR | R1 | 20240510 | 9MH2 | T3 |

Fig. 3. Examples of Secondary Hash Structures.

### *C. Optimal Storage Mechanisms for AI IDs*

To ensure authenticity, accountability, and traceability of AI systems, while preserving trade secrets, we recommend a blockchain-based AI identity system paired with a Zero-Knowledge Proof (ZKP) verification layer. This architecture ensures AI systems remain verifiable and compliant even as they operate autonomously in diverse or high-risk environments. Decentralized storage and verifiable provenance are key features of persistent system identity, enabling long-term accountability across distributed digital ecosystems. In this framework, blockchain-based registration is not employed to eliminate trust or automate approval decisions. Instead, it functions as an append-only, tamper-resistant log that preserves the historical state of AI registrations and enables verification across organizational and jurisdictional boundaries. The governance authority retains decision-making power, while the registry ensures that those decisions cannot be silently altered or retroactively rewritten. The framework does not require a public or permissionless blockchain. Depending on regulatory context, a permissioned ledger, consortium-managed registry, or other append-only infrastructure with equivalent audit guarantees may be sufficient. The essential requirement is not decentralization per se, but durable, verifiable preservation of identity records over time.
Zero-knowledge proof is not presented as a universal requirement for AI identification. Their use is justified primarily in deployment settings where AI systems operate across organizational boundaries, execute autonomously, or must prove registration status without continuous access to a central authority. In more tightly controlled environments, alternative verification mechanisms may be appropriate.

1) *Blockchain-Based AI ID:* The blockchain functions as a decentralized, tamper-resistant registry for logging the digital credentials of AI systems. Each entry includes the primary hash, secondary hash, ZKP anchor, and version metadata. This registry enables users, regulators, and independent verifiers to confirm the authenticity of any AI instance without relying on a centralized authority or exposing sensitive internal details. Combined with ZKPs, AI agents can prove registration status without exposing model weights, architecture, or proprietary IP. That enables transparent yet private AI attestation.
2) *Blockchain Registration Protocol:* We propose a five-step protocol for a blockchain-based AI ID system:
   - Step 1: Smart Contract Deployment—The registration authority publishes a registration smart contract to manage AI entries. It stores primary hash, secondary hash, and the ZKP anchor.
   - Step 2: Off-Chain Metadata Preparation—AI developers prepare a signed metadata bundle that satisfies primary and secondary hash requirements. The AI product company does not share AI-trained weights with the authority.
   - Step 3: Registration—The registration authority verifies submitted information, then digitally signs and issues an AI credential. This credential links the model fingerprint to its blockchain entry. An initial testing status of “U” (Undecided) is also recorded and updated post-testing.
   - Step 4: On-Chain Logging—Once validated, the primary hash, ZKP anchors, secondary hashes, developer signatures, and testing status are immutably recorded on-chain. A status of “P” (Pass) is required before deployment in regulated domains; failure results in a “F” (Fail) or “X” (Retired) flag, visible to validators.
   - Step 5: Credential Embedding and Distribution—The final AI credential, including the primary hash, secondary hash, and ZKP anchor, is embedded into each AI instance by the developer. This enables the AI system to prove its identity at governance checkpoints by generating a valid ZKP without revealing proprietary internals.

This registration workflow aligns with sustainable digital governance principles by ensuring that AI systems entering operational environments are consistently identifiable, auditable, and compliant throughout their lifecycle.

### *D. Registration-As-a-Precondition Assumption, Implementation Constraints and Operational Parameters*

The above framework assumes that participation in regulated digital environments requires proof of valid AI registration. The framework does not assume the elimination of institutional authority. Initial registration, approval, and policy enforcement may remain centralized or delegated to designated governance bodies, regulators, or certification entities. Under this assumption, AI systems are not permitted to execute unless they can demonstrate a registered identity via cryptographic verification. Consequently, adversarial attempts to deploy stolen or minimally modified model weights without re-registration do not evade governance controls: any alteration to the weight configuration invalidates the original hash commitment, preventing successful registry proof. Unregistered or unverifiable models are therefore flagged as suspicious, subject to monitoring, and may be disabled or reported by platform operators or regulatory authorities. This design does not seek to prevent all illicit AI use in unregulated contexts, but rather establishes enforceable identity checkpoints within enterprise, platform, and regulatory infrastructures, where governance-enforceable traceability is practically exercised.

In practice, adoption of AI identification infrastructure is likely to proceed incrementally rather than universally and can be understood as occurring in progressive stages of organizational uptake:

- Initial adoption is most plausible in regulated, high-risk, or platform-mediated environments, where identity verification can be embedded as a condition of access or operation.
- Intermediate adoption may occur as compliance tooling matures and organizations integrate AI identification into enterprise architecture and governance workflows.
- Broader diffusion may follow over time as interoperability benefits become apparent and identity continuity reduces duplication of certification and audit effort.

A natural question is why such a registry cannot be implemented as a manual database or spreadsheet maintained by a trusted authority. While centralized records may suffice for initial registration, they are poorly suited to long-term governance objectives that require immutability, historical auditability, and cross-organizational verification. Mutable records cannot reliably demonstrate that an AI system operating today corresponds to the same system that was approved previously, nor can they support third-party verification without continuous trust in the record keeper. While the proposed framework is blockchain-agnostic at the conceptual level, its practical deployment assumes a permissioned or consortium-based ledger, or a Layer-2 rollup anchored to a public blockchain, rather than a fully open public chain. This design choice reflects governance, cost, and compliance considerations typical of regulated enterprise and public-sector environments. On-chain transactions are limited to compact metadata—including cryptographic hashes, ZKP anchors, status flags, and digital signatures—rather than full model artifacts, resulting in payload sizes on the order of kilobytes per registration event. Cryptographic hashing costs are incurred once per stable model version and are negligible relative to the computational cost of model training or large-scale inference. Zero-knowledge proof circuits are intentionally scoped to proving possession of registered cryptographic commitments rather than full in-circuit evaluation of model weights or inference behavior, reflecting current scalability limits of ZKP systems for large models. Accordingly, the framework does not assume universal or continuous ZKP verification, but rather governance-defined checkpoints where identity assurance is operationally necessary. Operational costs associated with registration and verification are expected to be borne primarily by model developers and platform operators, consistent with existing software certification and compliance regimes. These costs are modest relative to training and deployment expenditures and are concentrated at version boundaries rather than during routine inference. From a compliance perspective, the registry is designed to minimize the storage of personal or sensitive information, supporting data minimization and auditability requirements under regulatory regimes such as GDPR. Collectively, these operational parameters ensure that the framework remains technically feasible and deployable within realistic governance and infrastructure constraints.

## *E. Governance Checkpoint Verification of AI Identity*

The framework employs zero-knowledge proof (ZKP)-based mechanisms to enable governance checkpoint verification of AI identity without disclosing proprietary model internals. Checkpoint ZKP verification confirms correspondence between a deployed system and a registered cryptographic commitment; it does not attest to inference correctness, safety properties, or compliance with downstream policy requirements. Through this approach, AI systems can demonstrate that they correspond to a registered cryptographic identity while preserving confidentiality of weights, architecture, and training data. Checkpoint verification supports ongoing identity assurance that deployed AI systems remain aligned with their registered identities.
The checkpoint verification process consists of following four coordinated phases. Together with the registration workflow described above, these phases complete the process of assigning, registering, and validating an AI ID. Figure 4 illustrates the end-to-end process.

- Step 6: Circuit Setup. Validation rules, such as acceptable drift thresholds or registration status requirements, are encoded into a ZKP circuit during the registration phase.
- Step 7: Verifier Challenges. At governance checkpoints, external verifiers, including platform operators or organizational control systems, query the registry and issue an identity challenge to the AI instance.
- Step 8: Proof Generation. The AI system generates a zero-knowledge proof using its embedded model state as a private witness.
- Step 9: Validation and Response. The verifier checks the proof against the on-chain ZKP anchor and associated AI-ID. Successful verification confirms that the AI instance is registered and has not been tampered with; failure triggers escalation or enforcement actions as defined by the deployment environment.

Governance checkpoint verification strengthens lifecycle governance by enabling continuous identity assurance across deployment contexts without requiring disclosure of sensitive intellectual property. It is best understood as a governance-defined checkpoint rather than a continuous cryptographic requirement, and may be selectively enabled based on deployment risk, autonomy, and regulatory context. The threat model and enforcement assumptions governing such verification are detailed in the following subsection.

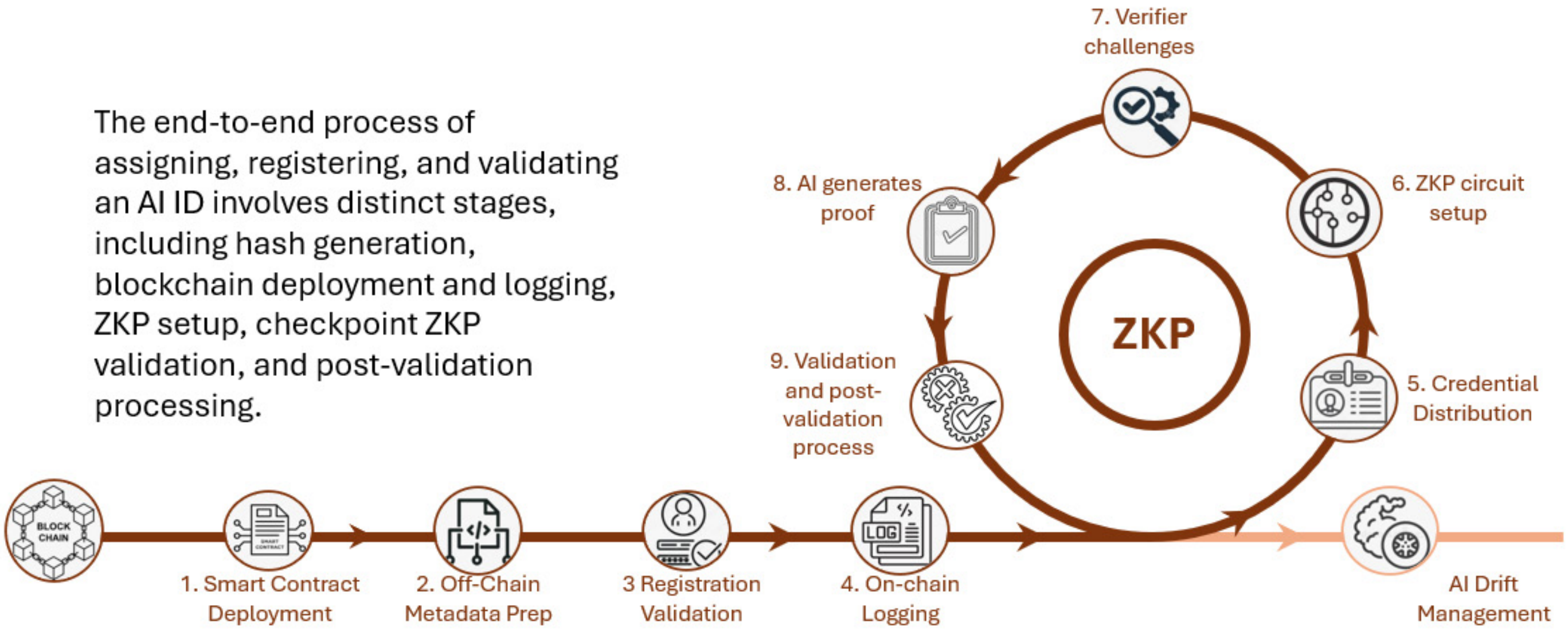

Fig. 4. End-to-end AI registration and validation process.

## *F. Threat Model and Enforcement Scope*

The proposed framework assumes a threat model in which AI model weights may be copied, modified, or redeployed without authorization, and in which adversaries may attempt to operate unregistered or tampered AI systems. The framework is therefore designed to support detection, attribution, and governance enforcement rather than to prevent all forms of illicit AI use at the technical level.

Enforcement is assumed to occur within regulated or enterprise-controlled execution environments, such as cloud platforms, application marketplaces, and organizational IT infrastructures, where AI systems are required to demonstrate valid registration and identity verification prior to operation. Within such environments, models that fail to produce a valid registry proof are treated as unregistered, flagged as suspicious, and subject to monitoring, restriction, or deactivation in accordance with applicable governance policies.

The framework does not aim to prove inference correctness, eliminate malicious intent, or prevent offline or illicit AI deployment in unregulated contexts that bypass institutional controls. These scenarios are considered outside the scope of this work. Instead, feasibility is discussed at a high level to clarify what is technically plausible under current toolchains and governance constraints. Cryptographic hashing is incurred once per stable model version and is negligible relative to the cost of training or large-scale inference. Zero-knowledge proofs are intentionally scoped to proof-of-possession of registered commitments rather than full in-circuit evaluation of model weights or inference behavior, reflecting current scalability limits in ZKP/ZKML systems for large models.

While more comprehensive technical containment mechanisms exist, the framework prioritizes practical enforceability and regulatory realism over universal technical containment.

### *G. Illustrative Feasibility and Performance Considerations*

The framework does not present empirical benchmarks for hashing, zero-knowledge proof generation, or similarity scoring. Instead, feasibility is discussed at a high level to clarify what is technically plausible under current tool chains and governance constraints, as outlined in the preceding section. Cryptographic hashing is incurred once per stable model version and is negligible relative to the cost of training or large-scale inference. Zero-knowledge proofs are intentionally scoped to proof-of-possession of registered commitments rather than full in-circuit evaluation of model weights or inference behavior, reflecting current scalability limits in ZKP/ZKML systems for large models.

In addition, similarity-based screening is presented as a governance-oriented mechanism for detecting structural divergence rather than a validated proxy for semantic drift, task-level performance change, or safety. Prior work on file-and artifact-level similarity measures, including LZJD, supports the basic plausibility of scalable structural comparison for large binary objects, while recent ZKML toolchains and prototypes illustrate both opportunities and present constraints for deploying ZKP-based verification in ML contexts. All performance-related considerations in this work are therefore intended as order-of-magnitude reasoning to establish architectural plausibility, not as claims of system-level efficiency or optimality.

## VIII. Post-Deployment Model Drift and Tampering Detection

A persistent governance challenge lies in verifying that deployed AI systems remain aligned with their registered identities over time. Once an AI model is registered, its learned parameter configuration serves as the anchor for identity verification. In practice, however, models may undergo fine-tuning, incremental retraining, optimization, or maintenance updates that introduce structural changes to the underlying weight configuration. Requiring full re-registration for every such modification would be operationally burdensome and could undermine the stability of regulatory oversight. Conceptually, this approach is analogous to a driver’s license photo: while an individual’s appearance may change over time, identity remains anchored to a stable, verifiable reference, and only substantial divergence necessitates re-issuance.

To balance identity continuity with technical realism, the proposed framework distinguishes between minor, governance-acceptable modifications and substantial divergence that warrants re-registration. This distinction is operationalized through similarity-based screening mechanisms that compare a model’s current weight configuration to its registered anchor state. In this context, similarity assessment is not intended to establish semantic or functional equivalence, but to provide a scalable and reproducible signal for governance decision-making regarding identity continuity.

In this framework, the Lempel–Ziv Jaccard Distance (LZJD) is proposed as a practical file-level similarity metric for post-deployment screening. LZJD operates by extracting sets of unique substrings from serialized byte streams using Lempel–Ziv parsing and computing the Jaccard distance between the resulting sets. The metric produces a normalized score between 0 and 1, where lower values indicate greater structural similarity. Prior work has shown LZJD to be computationally efficient and effective at detecting structural divergence across large binary artifacts, making it suitable for large-scale model comparison in governance contexts [34].

Importantly, LZJD is employed as a conservative screening mechanism rather than a definitive proof of model equivalence. Accordingly, LZJD is not proposed as a validated measure of semantic drift, task performance change, or model safety, but solely as a reproducible structural signal to support governance decisions regarding identity continuity. Like other file-based similarity measures, LZJD is sensitive to changes in serialization format, numerical precision, checkpoint layout, quantization, pruning, or compression, even when high-level model behavior is largely preserved. For this reason, the metric is not used to infer behavioral correctness, safety, or task-level performance. Instead, it functions as a reproducible signal for detecting substantial structural deviation relative to a registered anchor configuration. This distinction is intentional. Structural divergence refers to detectable changes in a model's serialized parameter configuration, whereas semantic drift refers to changes in task behavior, output distributions, or performance characteristics. The former is necessary for governance-enforceable identity continuity, while the latter requires domain-specific evaluation methods that fall outside the scope of this work.

In this framework, similarity scoring is treated as a governance signal rather than a scientific measurement, intended to trigger review or re-registration decisions rather than to quantify model semantics, safety, or task performance. Under the proposed framework, acceptable similarity thresholds are governance-defined rather than scientifically fixed but are defined by regulators or organizational governance bodies based on deployment context, model class, and risk category. Thresholds may therefore vary across sectors and applications. When a model's similarity score remains within the approved threshold, identity continuity is preserved and re-registration is not required. When the score exceeds the threshold, the model is treated as drifted, triggering governance actions such as re-registration, additional review, or deployment restriction. The framework does not prescribe specific numeric thresholds, but provides an auditable mechanism through which such thresholds can be consistently applied and enforced.

To support accountability and prevent circumvention, regulators or oversight bodies may require attestations that similarity assessments were performed on the currently deployed model rather than on outdated or substituted artifacts. Such attestations may be supported through cryptographic hashes or zero-knowledge proof-backed claims that bind the similarity calculation to the relevant model versions without disclosing proprietary weights.

While LZJD is presented as one practical and scalable option, the framework is deliberately metric-agnostic and does not depend on any single similarity function. Alternative or complementary approaches, including parameter-wise distance measures, layer-wise summaries, or architecture-aware comparison techniques, may be adopted depending on regulatory preference or technical requirements. In practice, similarity-based screening may be implemented as a first-tier filter, with more detailed evaluation procedures triggered only when deviation exceeds governance-defined thresholds. Governance authorities or organizations may therefore substitute alternative structural comparison techniques without altering the underlying identification architecture.

By enabling scalable detection of structural divergence without mandating full re-registration for every minor update, the proposed drift detection approach supports persistent system identity. It preserves lifecycle accountability while avoiding unnecessary retraining, recertification, and audit duplication, thereby contributing to more efficient allocation of computational, organizational, and regulatory resources over time.

## IX. Limitations

Several limitations of this work merit acknowledgment:

- First, the pace of AI development is rapid, and while we conducted an extensive literature review, it is possible that recent or emerging contributions may have been overlooked. This paper reflects a current understanding rather than an exhaustive survey.
- Second, the framework presented adopts specific technologies, for example, cryptographic hashing, blockchain registration, ZKP-based validation, and LZJD drift detection, to illustrate feasibility. These are illustrative rather than prescriptive; in practice, alternative implementations may be equally viable depending on technical or regulatory context. In particular, similarity-based screening is not intended to capture semantic equivalence or safety properties, and must be interpreted strictly within a governance and identity-continuity context. The inclusion of zero-knowledge proofs reflects one approach to privacy-preserving verification; in some deployment contexts, simpler attestation mechanisms may achieve equivalent governance objectives.
- Third, the proposal assumes a cooperative environment in which developers and regulators participate in good faith. In adversarial settings or under weak enforcement, compliance may be limited, which could require regulators to conduct site visits and implement other measures.
- Fourth, the framework is not empirically validated in this study. While specific experimental evaluations, e.g., benchmarking ZKP proof generation times, measuring hashing overhead for large-scale models, and evaluating LZJD sensitivity under quantization, pruning, or fine-tuning, are necessary for operational deployment, they fall outside the scope of this conceptual and governance-focused contribution.
- Finally, adoption of AI identification infrastructure depends on institutional incentives, regulatory mandates, and organizational capacity. Absent external requirements or coordination mechanisms, voluntary uptake may be limited, particularly for smaller organizations. The framework therefore presumes deployment in regulated or high-risk contexts where durable accountability provides clear governance and economic value, rather than universal adoption across all AI use cases.

## X. Future Validation Items

Future research should prioritize empirical validation of the specific feasibility questions raised in this study, including controlled experiments evaluating similarity-based drift detection under quantization, pruning, and fine-tuning, benchmarking zero-knowledge proof generation times across model scales, and measuring hashing overhead for large-scale models. This includes controlled experiments evaluating similarity-based drift detection under common model modifications, benchmarking zero-knowledge proof generation times across model scales, and analyzing economic and organizational incentives for adoption under different regulatory regimes. Such studies are necessary to support operational deployment, but are analytically distinct from the conceptual and governance contribution advanced in this paper.

## XI. Summary

This paper proposes an integrated framework for AI identification that bridges technical feasibility with regulatory enforceability. By anchoring AI identity in its learned weight configuration, generating cryptographic identifiers, and storing them immutably on a blockchain registry, the approach enables transparent, tamper-resistant provenance. Governance checkpoint validation is supported through zero-knowledge proofs, while post-deployment drift is monitored using LZJD-based similarity scoring.

Together, these mechanisms establish a lifecycle-oriented model for AI traceability that supports public oversight without compromising proprietary information. Such lifecycle traceability is increasingly essential for persistent system identity, where long-term accountability, transparency, and risk management depend on the ability to verify and monitor AI systems across their operational lifespan. By separating governance authority from identity record durability, the proposed framework clarifies why robust registries are necessary even in centralized approval regimes. By aligning governance requirements with operational incentives, the framework illustrates how AI identification can function as enabling infrastructure rather than a purely regulatory burden. While not exhaustive, the framework offers a scalable and policy-aligned foundation for future AI registration and compliance systems, and when integrated with enterprise architecture processes such as AIDAF, can serve as a core building block of sustainable digital governance and transformation. By embedding verifiable AI identity into broader governance, architecture, and digital transformation workflows, the framework contributes to the development of responsible, resilient, and sustainable AI-enabled enterprises and public-sector systems.
Beyond mechanisms of control and accountability, the sustainability contribution of AI Identification lies in its support for long-term institutional resilience. By enabling persistent system identity, the framework reduces retraining, re-certification, and re-audits, thereby supporting more efficient allocation of computational, organizational, and regulatory resources over time. AI identification thereby contributes to sustainability not only by improving oversight, but by strengthening the durability, continuity, and scalability of digital governance arrangements as AI systems evolve.

The work reported herein was supported by the National Science Foundation (NSF) (Award #2246920). Any opinions, findings, and conclusions or recommendations expressed in this material are those of the authors and do not necessarily reflect the views of the NSF.

## Appendix

This appendix provides a high-level, non-normative pseudocode illustration of the AI identification workflow discussed in the main text. The purpose is to clarify conceptual sequencing rather than to prescribe an implementation or performance characteristics.

**Algorithm 1:** Illustrative AI Registration and Verification Procedure

**Input:** Model parameters $W$, model metadata $M$, registry namespace $N$.
**Output:** Registered AI identifier $AI_ID$.
**Steps:**

1) Compute a structural fingerprint of the model parameters:

$$F \leftarrow \text{Fingerprint}(W)$$

2) Compute a model-level cryptographic hash from the fingerprint:

$$H_1 \leftarrow \text{Hash}(F)$$

3) Compute a namespaced registry commitment using organizational or registry context:

$$H_2 \leftarrow \text{Hash}(H_1 \,||\, N)$$

4) Register the namespaced commitment and associated metadata reference:

$$\text{Register}(H_2, M)$$

5) *(Optional)* Generate a proof-of-possession of the registered commitment:

$$\pi \leftarrow \text{Prove}(W, H_2)$$

6) At governance-defined deployment or audit checkpoints, verify identity:

$$\text{Verify}(\pi, H_2)$$

*Note:* This pseudocode is provided solely for conceptual illustration. It is non-normative and does not represent a reference implementation, performance benchmark, or mandated technical configuration.
The procedure illustrates how AI identification is bound to a model's internal configuration rather than its documentation or observed behavior. Registration occurs at version boundaries, while verification is performed selectively at governance-defined checkpoints. The workflow is intentionally modular and technology-agnostic, allowing alternative fingerprinting, hashing, or proof mechanisms to be substituted without altering the underlying identity architecture.